# Simple method for mid-infrared optical frequency comb generation with dynamic offset frequency tuning

Mikhail Roiz[1,*], Krishna Kumar[1], Juho Karhu[1,2] & Markku Vainio[1,3]

*1 Department of Chemistry, University of Helsinki, Helsinki, Finland*

*2 Metrology Research Institute, Aalto University, Espoo, Finland*

*3 Photonics Laboratory, Physics Unit, Tampere University, Tampere, Finland*

e-mail: markku.vainio@helsinki.fi ; mikhail.roiz@helsinki.fi

**Abstract**

We present a simple method for fully-stabilized mid-infrared optical frequency comb generation based on single-pass femtosecond optical parametric generation that is seeded by a continuous-wave laser. We have implemented the method in a periodically poled lithium niobate crystal that produces a frequency comb tunable across 3325 – 4000 nm (2380 - 3030 cm$^{-1}$). The method generates the mid-infrared (idler) comb with known and stabilized Carrier-Envelope Offset (CEO) frequency without the need to directly detect it. The idler CEO is continuously tunable for almost half of the repetition rate and can be modulated, while maintaining its central frequency. Together with the high output power (up to 700 mW) and low intensity noise (0.018% integrated in 10 Hz - 2 MHz bandwidth) this makes the demonstrated mid-infrared frequency comb promising for many applications such as high-precision molecular spectroscopy, frequency metrology and high harmonic generation.

## I. Introduction

Since their first development about 20 years ago, Optical Frequency Combs (OFC) have advanced science and technology in many different ways. By creating a phase-coherent link between optical and radio frequencies[1], OFCs have enabled accurate measurements of optical frequencies with a simple and elegant implementation leading to a rapid progress in the development of new optical frequency standards[2]. The establishment of OFCs has only become possible thanks to the invention of novel techniques for precise measurement and control of Carrier-Envelope Offset (CEO) frequency – a key parameter for ensuring pulse-to-pulse coherence in femtosecond Mode Locked Lasers (MLL)[3, 4]. Consequently, OFCs have found applications in a variety of different fields[5-7] including optical atomic clocks[8], optical arbitrary waveform generators[9], precise ranging, telecommunications and molecular spectroscopy[10, 11].

Despite the fact that MLL technology opened up a new path to high-precision molecular spectroscopy, a major challenge has been to access the Mid-Infrared (MIR) spectral region, where a variety of strong molecular fingerprints can be measured[10, 12]. Although many alternative techniques for MIR OFC generation were proposed, including direct MIR OFC generation by quantum cascade lasers[13], in general it requires the nonlinear frequency conversion processes to be involved. The two most common methods for MIR OFC generation based on nonlinear conversion are Difference Frequency Generation (DFG) and Synchronously-Pumped Optical Parametric Oscillation (SPOPO). In most cases, DFG requires either two fully stabilized Near-Infrared (NIR) OFCs for the input[14], or one NIR OFC and its extended version via Raman downshifted soliton in supercontinuum (SC) generation[15-18]. While the second option of DFG allows one to use a single NIR comb for the pump and signal, this implementation leads to the cancelation of CEO (it is always 0 for the MIR comb), and thus the CEO cannot be easily changed. SPOPO does not make the CEO tuning easier, since for a singly resonant SPOPO one needs to precisely control the pump CEO and the cavity length at the same time to tune the MIR comb CEO, and determination of the exact CEO frequency requires an additional measurement setup[12, 19, 20]. In doubly resonant degenerate SPOPO[21] the CEO tuning can be performed just by tuning the CEO of the pump laser, but this scheme implies the random choice between two possible CEO values for the MIR comb[20, 22-24], complicating its CEO determination. In addition, SPOPOs generally require precise cavity locking[25] and careful engineering of the group delay dispersion[22].

In the present work, we demonstrate a simple approach for the generation of fully stabilized MIR OFCs with CEO tuning and modulation. The approach is based on single-pass femtosecond Optical Parametric Generation (OPG) seeded by a continuous-wave (CW) laser, and it can be applied to any OPG. Importantly, OPG on its own does not directly lead to the generation of stable OFCs for the signal and idler, since the process starts from noise, and thus the CEO is random for two subsequent signal and idler pulses. In order to fully stabilize the output MIR comb, we use a CW laser phase-locked to the pump laser to seed the signal comb. Several important advantages arise here compared to DFG and SPOPO. First, the idler CEO is defined by the radio-frequency (RF) local oscillator (LO) used for phase-locking of the seed laser, which means that there is no need to measure the CEO, since it is always known. Second, the CEO of the MIR comb is continuously tunable by simply changing the RF LO frequency. It makes the system highly versatile combined with easily tunable repetition rate thanks to the cavity free design. Third, the idler CEO can also be modulated at relatively high frequencies precisely maintaining its CEO central frequency. Moreover, the method is inherently free of any optical self-referencing techniques like f-2f[4, 5] and 2f-3f[26] interferometry meaning that the CEO of the pump MLL can be left free running. We believe that the presented simple method for the generation of fully-stabilized MIR OFCs has major advantages compared to the existing methods, especially in applications that require precise control of the CEO and repetition rate - such as cavity enhanced spectroscopy[27], dual-comb spectroscopy[28-30], comb-assisted spectroscopy[31, 32] and high harmonic generation[33, 34]. Additionally, the setup demonstrated here can be used as a pump source for another OPG, DFG or SPOPO[35], since it has a large wavelength tuning range and high output power. The high output power is also vital for certain spectroscopic methods including OFC photoacoustic spectroscopy[36, 37] and background-free absorption spectroscopy[38]. In this article, we demonstrate the proof of concept implementation of the method, including rigorous characterization of the system.

## II. Principle

Usually, in OPG, high-energy pulses are used to pump a nonlinear crystal, which in non-degenerate case leads to the generation of signal and idler pulses. The generation of the new pulses is governed by the quantum noise amplification. This makes the process inefficient and leads to loss of pulse-to-pulse coherence of the produced signal and idler pulse trains. The situation changes if the OPG is seeded by an additional light source in the signal or idler spectral region, which is similar to Optical Parametric Amplifiers (OPA). First, the seeding may lead to a significant reduction of the OPG threshold[39]. Second, it improves the pulse-to-pulse coherence provided that the seeding source is coherent[40]. Third, it reduces the relative intensity noise (RIN) of the system[41], which is vital for many spectroscopy applications. In addition, it was demonstrated that two OPG setups seeded by the same CW seed laser produce mutually coherent output NIR combs (signal), which is useful for dual-comb spectroscopy[42]. Despite all the benefits, to the best of our knowledge, the possibility of using CW-seeded OPG to produce a fully-stabilized MIR OFC source has not been demonstrated before.

Numerous OPG and OPA experiments have been reported using different nonlinear materials including BBO[43], PPKTP[44], GaAs[45], LiTaO$_3$[46]. Of particular interest is the OPG in MgO doped periodically poled lithium niobate (MgO:PPLN) because this crystal has a high nonlinear coefficient, high damage threshold and large transparency range that extends to the 3 μm CH-stretching vibrational region, which is useful for molecular spectroscopy. Importantly, when the pump source has a pulse duration in the range of hundreds of femtoseconds or shorter, the interaction length may become an issue. Due to the difference in group velocities between the pump, signal and idler pulses, the interaction length can be so short that the OPG threshold is never reached without the risk of damaging the crystal. Fortunately, several research groups demonstrated that MgO:PPLN supports OPG where the interaction length can be as long as 50 mm for femtosecond pump pulses leading to the conversion efficiencies over 50% with low threshold[47, 48]. A distinctive feature of this particular case is the convenient pump wavelength of 1030 – 1064 nm that coincides with the mature technology of high power Nd:YAG solid state and Yb-doped fiber MLLs. When launched into the MgO:PPLN in the given spectral range, the high-energy pump generates the signal and idler pulses, whose group velocities have opposite signs and equal absolute values leading to a so-called pulse trapping effect[49].

In general, each optical frequency component $\nu_n$ of a frequency comb can be represented as a combination of two radio frequencies using the following formula[5]:

$$\nu_n = f_{CEO} + nf_r,$$

where $f_r$ is the repetition rate, $f_{CEO}$ is the CEO frequency, and $n$ is the mode number of the given optical frequency component (comb tooth). We will use *CEO* instead of $f_{CEO}$ further in the text. As usual, $f_r$ is directly detectable using a fast photodiode, and thus easy to stabilize. On the other hand, CEO cannot be directly detected without an additional measurements setup, such as an f-2f interferometer[4]. Luckily, in OPG there is a way to determine, stabilize and freely tune the idler CEO without directly detecting it.

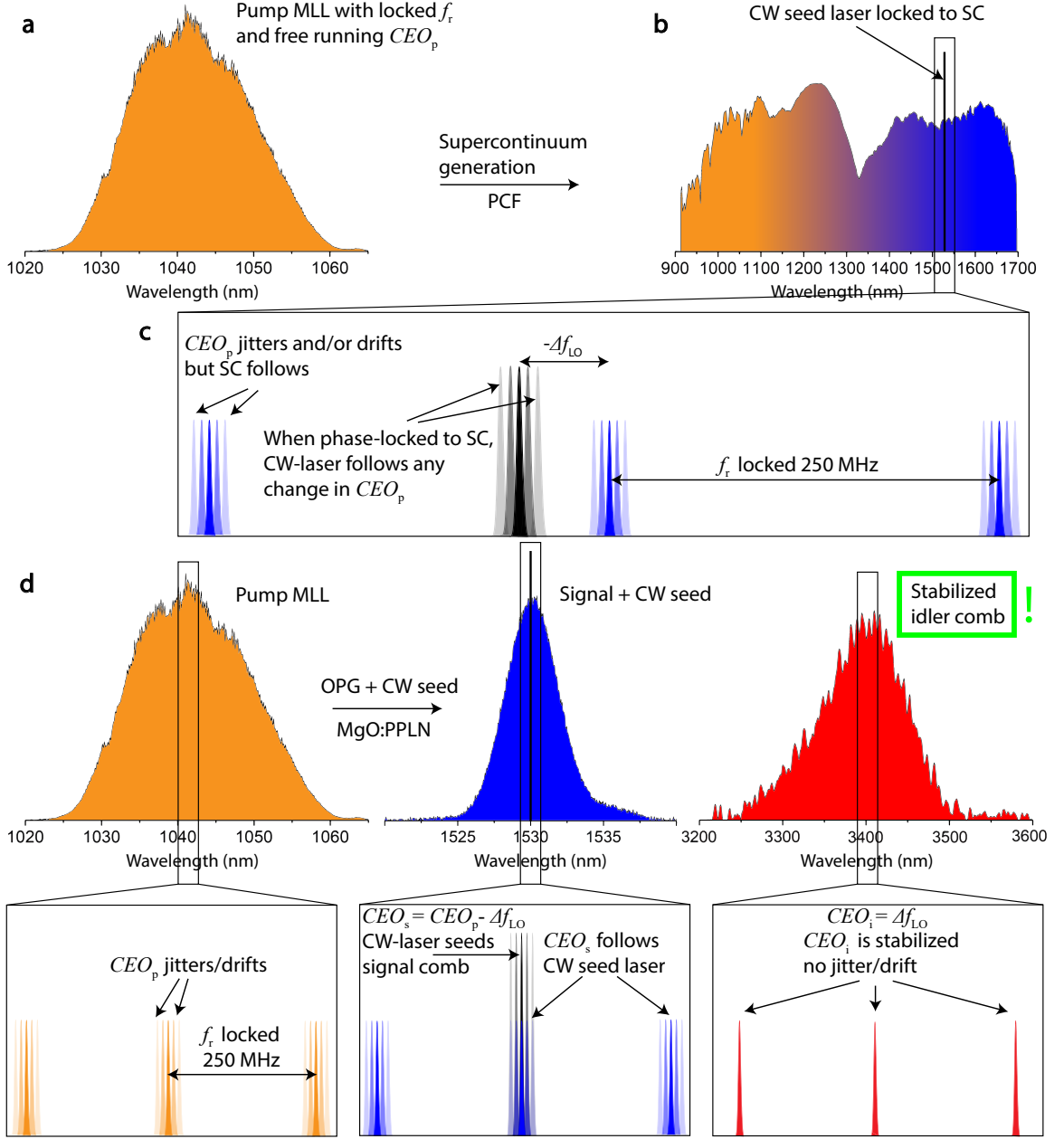

**Fig. 1** Schematic of the CW seeded OPG for stabilized MIR OFC generation. (a) Pump envelope spectrum. (b) SC envelope spectrum combined with CW seed laser. (c) CW seed laser (black line) locked to SC (blue lines) in the state corresponding to Eq. (2) – $CEO_s = CEO_p - \Delta f_{LO}$. (d) Generated signal and idler envelope spectra using CW seeded OPG (above) and the schematic of the corresponding laser modes (below).

Considering the principle of conservation of energy, the CEO equation for the OPG process can be written as follows:

$$CEO_p = CEO_s + CEO_i, \quad (1)$$

where $CEO_p$, $CEO_s$, $CEO_i$ are the offset frequencies of the pump, signal and idler combs, respectively. In regular OPG without seeding, even if the $CEO_p$ is stabilized, $CEO_s$ and $CEO_i$ are random. On the contrary, if one seeds the signal comb with a coherent source, the $CEO_s$ follows the seed source, which consequently leads to a change in $CEO_i$ to satisfy Eq. (1). We can use this feature in our favor to stabilize the $CEO_i$. Let us first assume that $CEO_s = CEO_p$; according to Eq. (1) it leads to $CEO_i = 0$, which is the case of DFG mentioned above[15, 16]. Here it does not matter if the $CEO_p$ is stabilized or not because the $CEO_i$ will always remain zero thanks to Eq. (1). Next, let us set the $CEO_s$ in a different way:

$$CEO_s = CEO_p - \Delta f \text{ or} \qquad (2)$$

$$CEO_s = CEO_p + \Delta f, \qquad (3)$$

where $\Delta f$ is a frequency offset; in this case we have two states: $CEO_i = \Delta f$, or $CEO_i = -\Delta f$, respectively for Eq. (2) and Eq. (3), which are easy to distinguish in our setup. Hence, if one can precisely set $\Delta f$, it means that the $CEO_i$ is known and freely tunable.

In practice, the abovementioned $CEO_i$ stabilization can be performed in the four steps schematically shown in Fig. 1. First, the output beam of the pump MLL is split into two arms. Note that in order to produce fully stabilized MIR comb, repetition rate of the pump MLL should be stabilized ($CEO_p$ can be left free running). Second, the first arm is used to generate SC that reaches the CW seed laser wavelength (see "Supplementary note 1: Supercontinuum generation" for details). Third, the CW seed laser frequency is phase-locked to the generated SC. When the phase-locking is performed, the frequency offset $\Delta f$ is defined by an RF LO, which is why we denote $\Delta f = \Delta f_{LO}$; see "Supplementary note 2: Phase-locking" for more information. Since $\Delta f_{LO}$ can be easily changed, the $CEO_i$ is continuously tunable. On the last step, the second arm of the pump MLL and the phase-locked CW seed laser are combined in a nonlinear crystal and used to generate the signal and idler combs via seeded OPG.

By phase-locking the CW seed laser to the SC produced from the pump MLL, we transfer all the changes and fluctuations of the $CEO_p$ to $CEO_s$, which according to Eq. (1) makes the $CEO_i$ stable. Note that in Fig. 1 we only consider the case of Eq. (2) that corresponds to $CEO_i = \Delta f_{LO}$. The states in Eq. (2) and Eq. (3) can be distinguished in the phase-locking process by simply monitoring the behavior of the RF beat note between the CW seed laser and the SC. The state of Eq. (2) is realized when prior to locking the RF beat note frequency decreases while increasing the optical CW laser frequency (decreasing wavelength), if the opposite is true then the state corresponding to Eq. (3) is realized. All the envelope spectra in Fig. 1 (a,b,d) are examples of real measured spectra. The spectra are shown in the linear scale with arbitrary units except for the SC spectrum in Fig. 1 (b), which is shown in the decibel scale for clarity (see "Supplementary note 1: Supercontinuum generation" for details). We will call the wavelengths used in Fig. 1 (1530 nm signal and 3400 nm idler) "reference point" throughout the text.

### III. Experimental setup

Simplified schematic of our experimental setup is depicted in Fig. 2. Note that only essential components are shown in Fig. 2, and components with secondary importance such as half-wave plates, alignment mirrors, polarizing beam splitters etc. are omitted for simplicity. In our experimental setup we use a commercial Yb-doped fiber MLL (MenloSystems GmbH, Orange comb FC1000-250) as the pump source. It has 100 fs pulse duration Full-Width at Half-Maximum (FWHM), 250 MHz repetition rate locked to an RF source, <200 kHz comb tooth linewidth (FWHM), 10.5 W maximum output average power and free running $CEO_p$. For the SC generation we use an 80 cm PCF with two zero dispersion wavelengths (NKT Photonics, NL-PM-750). In order to generate the SC with the optical bandwidth shown in Fig. 1 (b), our PCF required 200 mW of average pump power with a coupling efficiency of 50% (so only 100 mW reaches the PCF). The SC is optically filtered using a diffraction grating at 1530 nm with a bandwidth of 5 nm (FWHM) and combined with the CW seed laser (also set to 1530 nm) for the phase-locking. See "Supplementary note 1: Supercontinuum generation" for more information. The CW seed laser is a commercial ECDL (Toptica Photonics, CTL 1550) that is locked to the SC using a PID controller (Toptica Photonics, mFALC 110) and an RF LO (Rohde & Schwarz, SME03). In addition, all the electronic instruments used in the experiments are referenced to a 10 MHz GPS-disciplined frequency reference (MenloSystems GmbH, GPS-8), which is not shown in the Fig. 2.

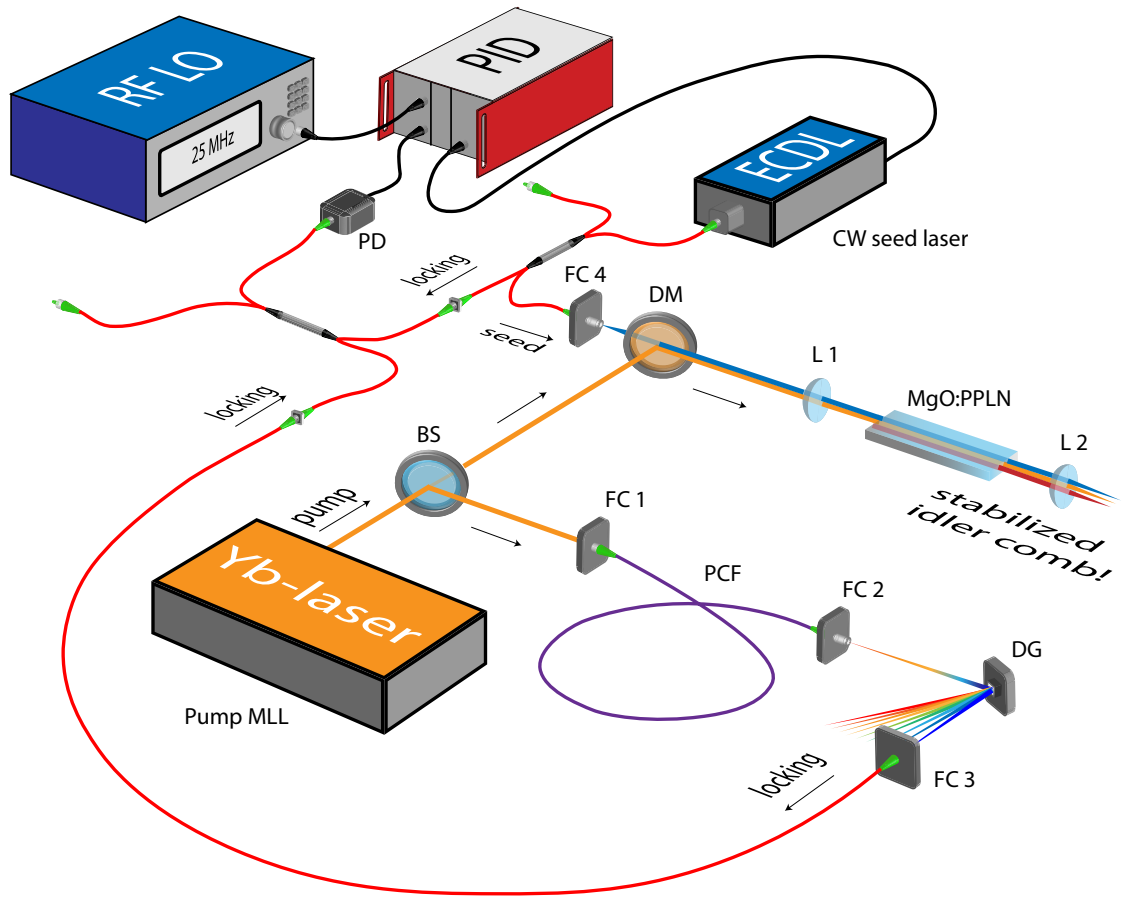

**Fig. 2** Simplified experimental setup of the CW seeded OPG. The pump MLL is split into two arms: One is used to produce SC in a PCF and another is used to pump the OPG in MgO:PPLN. The output of the ECDL is split into two arms for the seeding and phase-locking. The SC is optically filtered using DG and combined with the CW seed laser to produce an RF beat note on the PD for phase-locking. The phase-locking is performed using PID and RF LO; BS – beam splitter, FC – fiber collimator, DG – diffraction grating, DM – dichroic mirror, L – lens, PD – photodiode, PID - Proportional–Integral–Derivative controller, ECDL – External Cavity Diode Laser, PCF – Photonic Crystal Fiber.

The MLL is used to pump a 50 mm long 5% MgO doped PPLN fanout crystal (HC Photonics) with quasi phase-matching (QPM) periods of 26.5 – 32.5 µm. We use a focusing parameter of ξ = 2.17 close to an optimum 2.84 according to Boyd-Kleinman theory[50], which resulted in 84 µm waist diameter inside the MgO:PPLN crystal. The crystal is placed in an oven with the temperature set to 75 ºC to reduce the possibility of photorefractive damage. The operating wavelengths can be continuously tuned by translating the crystal in the laser beam path so that different QPM periods of the fanout structure are used. The seed CW laser beam is delivered to the OPG crystal via an optical fiber, followed by a fiber collimator for free space transfer through a dichroic mirror to the crystal. The seed laser waist size and position in the crystal are matched with those of the generated signal.

### IV. Results

**Operating wavelengths, thresholds and output powers.** Fig. 3 (a) shows the average pump powers required to reach the OPG threshold for different operating wavelengths without seeding (red) and with 5 mW of CW seed power (blue). Due to the limited tuning range of our seed laser (1510 – 1630 nm), we could not seed the OPG for signal wavelengths shorter than 1510 nm. At the reference point, the OPG requires 3.5 W of average pump power or 2.37 GW/cm$^2$ peak intensity to reach the threshold without CW seeding, and 2.4 W (1.63 GW/cm$^2$) with 5 mW CW seed power, which is about 32% threshold reduction. Using 5.4 W (3.66 GW/cm$^2$) of pump power and 5 mW of seed power at the reference point, the average idler power can be as high as 700 mW (1.5 W of average signal power) with an optical bandwidth of more than 100 nm (FWHM). The damage threshold of our crystal was experimentally determined to be 6.8 W (4.61 GW/cm$^2$), which led to the grey tracking effect.

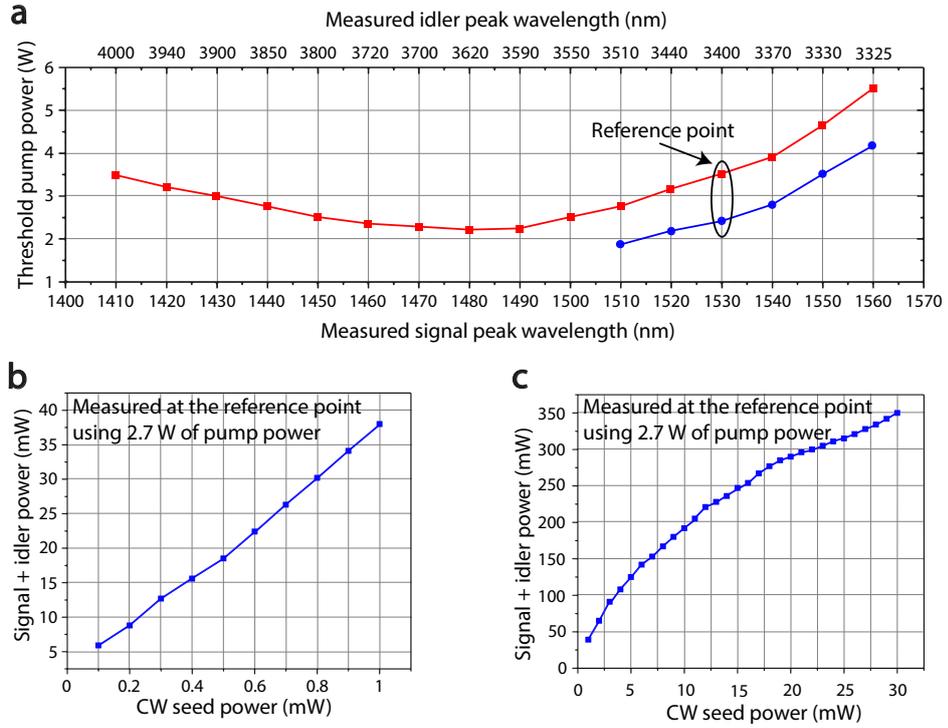

**Fig. 3** OPG power measurements. (a) Input average pump power vs. operating wavelengths for non-seeded OPG (red squares) and 5 mW CW seeded OPG (blue circles). Note that the threshold increases at short and long operating wavelengths because of the differences in group velocities between the pump, signal and idler pulses. (b), (c) Total converted OPG power (signal + idler, CW seed power excluded) at low and high CW seed powers, respectively, when the average pump power is fixed at 2.7 W (below the non-seeded OPG threshold).

The effect of CW seed power on the total converted OPG power (signal + idler) is demonstrated in Figs. 3 (b,c). In these measurements, the input average pump power was fixed at 2.7 W, which is below the threshold of non-seeded OPG. As can be seen in Fig. 3 (b), the OPG threshold is still reached even at 100 µW of CW seed power. In addition, one can significantly increase the OPG output power by simply using higher CW seed power (see Fig. 3 (c)). Nevertheless, for the measurements described further in the text, we used 5 mW of CW seed power, which is attainable with most semiconductor lasers at the telecom wavelengths.

**Idler CEO stability and tuning.** Let us start the discussion of $CEO_i$ stability from the CW seed laser phase-locking. Using the reference point, we first locked the seed laser to the state governed by Eq. (3) where $\Delta f = \Delta f_{LO} = 25$ MHz. Figure 4 (a) shows the corresponding RF beat note between the seed laser and the SC when the phase-locking is established. Next, we verified the phase-locking quality by measuring the double-sideband phase-noise of the beat note shown in Fig. 4 (a) using the phase-detector method[51]. The integrated phase noise (100 Hz – 2 MHz bandwidth) is as low as 16.5 mrad meaning that the CW seed laser is tightly locked to the SC. In addition, we performed frequency counting of the same RF beat note. The result can be seen in Fig. 4 (c) that shows the standard deviation of 34 mHz confirming high quality of the phase-locking. The noise in the frequency counting experiment averages out as the white noise, which is evident from the Allan deviation plot shown in Fig. 4 (d). It is worth emphasizing that if one wants to make sure that the $CEO_i$ is stabilized, the data shown in Fig. 4 (a-d) can be used as a reference. In order to prove this statement, we measured the $CEO_i$ and verified its stability, which we discuss next.

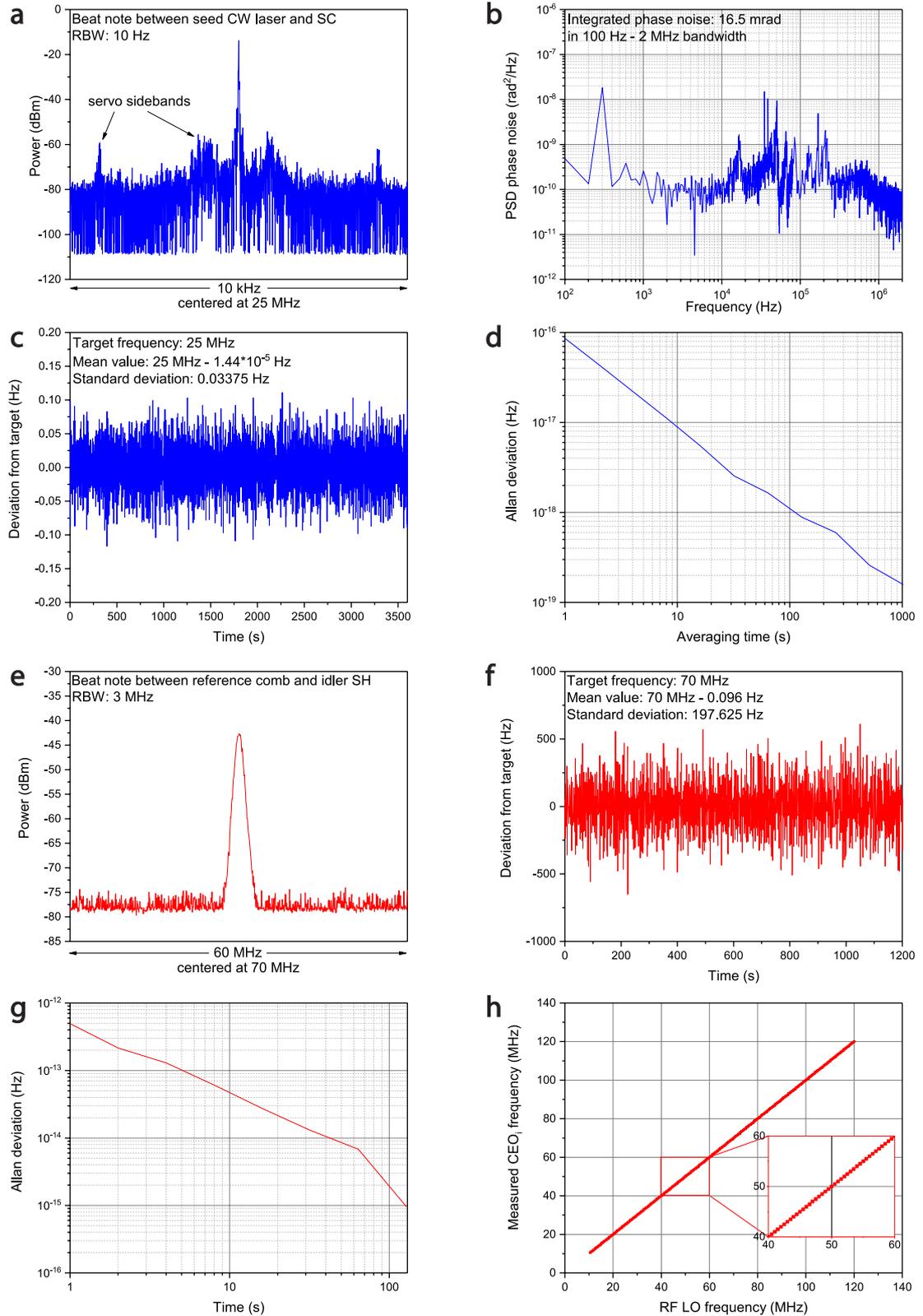

**Fig. 4 $CEO_i$ stability measurements and tuning.** (a) RF beat note of the CW seed laser locked to the SC. (b) Double-sideband phase-noise (PSD) of the RF beat note (shown in (a)) when the seed laser is locked to the SC. (c), (d) Frequency counting of the beat note shown in (a) and its Allan deviation, respectively. (e) RF beat note between the reference comb and the idler SH. (f), (g) Frequency counting of the RF beat note shown in **e** and its Allan deviation, respectively. (h) $CEO_i$ tuning demonstration; note that the $CEO_i$ is continuously tunable for almost 125 MHz, which is half of the idler repetition rate.

The most straightforward way to determine the CEO and measure its stability is f-2f interferometry. This would require an octave spanning SC spectrum generated directly in MIR, which is a challenging task. Instead, we emulated f-2f interferometry method by comparing the idler OFC to another OFC with a known and stabilized $CEO_{ref}$, which we will refer to as a reference comb. The reference comb is a commercial Er-doped fiber MLL (MenloSystems GmbH, Blue comb FC1500-250-WG) that has an inbuilt SC output spanning across 1050 – 2100 nm, with the $CEO_{ref}$ stabilized to 20 MHz. The reference comb does not directly overlap with the idler comb, hence we had to frequency double the idler comb using another MgO:PPLN crystal to produce an additional comb centered at 1700 nm. When frequency doubled, the CEO of the idler also doubles, meaning that the idler second harmonic (SH) offset frequency ($CEO_{iSH}$) is equal to twice the $CEO_i$. We combined the idler SH with the reference comb and set the repetition rates of both combs to 250 MHz. Using a simple delay line to overlap the optical pulses in time, we measured a single beat note with the following value:

$$f_{\text{beat}} = CEO_{\text{ref}} - CEO_{\text{iSH}} = 20 \text{ MHz} - (-2 * 25 \text{MHz}) = 70 \text{ MHz}.$$

One can see a stable beat note with the expected value of 70 MHz in Fig. 4 (e). It proves that the phase-locked CW seed laser transfers the instabilities of the free running $CEO_p$ to the signal comb ($CEO_s = CEO_p + \Delta f_{LO}$), thus making the $CEO_i$ stabilized according to Eq. (1). We used the same beat note for a frequency counting experiment that demonstrates the long term stability of $CEO_i$ (see Fig. 4 (f)). The noise in Fig. 4 (f) is limited by the mutual instability (timing jitter) between the pump comb (hence the idler comb and its SH) and the reference comb. The pump and reference combs are locked to two different RF signal generators that are both referenced to the same 10 MHz GPS-disciplined crystal oscillator. The specified relative instability of the reference oscillator is $5 \times 10^{-12}$ in 1 s.

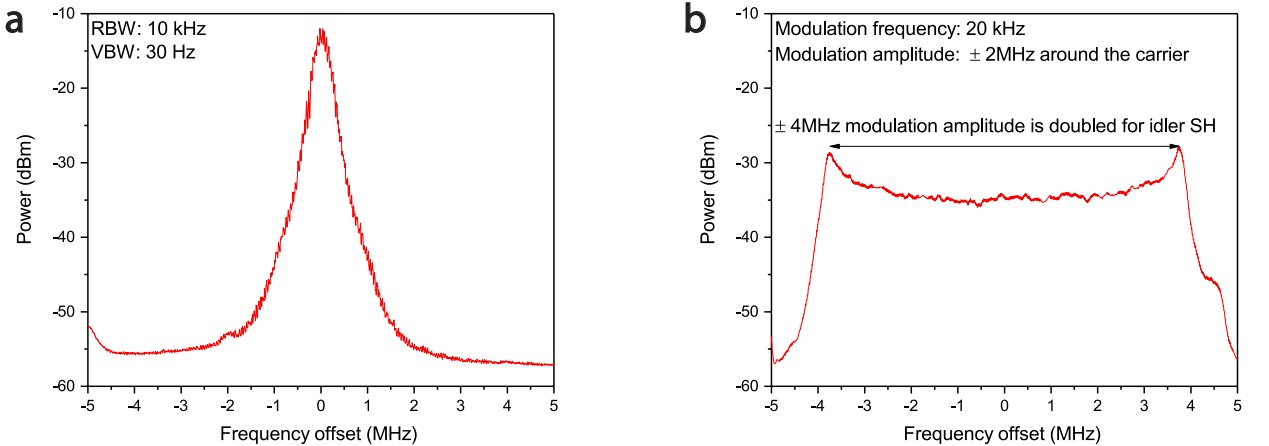

**Fig. 5 Idler CEO modulation experiment.** (a), (b) RF beat note between the reference comb and the idler SH without and with modulation, respectively.

In order to demonstrate the $CEO_i$ tuning, we swept the RF LO frequency from 10.5 MHz to 120 MHz with a step of 0.5 MHz and continuously measured the beat note frequency using an inbuilt peak finding function of the RF spectrum analyzer. The result can be seen in Fig. 4 (h). As evident from the figure, the idler CEO is continuously tunable for almost half of the idler repetition rate without any interruptions. For RF LO frequencies close to 125 MHz the phase-locking becomes unstable because the adjacent RF beat note from the next nearest comb tooth starts to overlap with the beat note under consideration. Nevertheless, when the 125 MHz point has passed, the $CEO_i$ can be further tuned without interruptions.

**Idler CEO modulation.** Next, we modulated the RF LO frequency $\Delta f_{LO}$ with a maximum modulation frequency of 20 kHz available in our RF LO. In addition, we chose the largest modulation amplitude of ±2 MHz that allowed for stable phase-locking without changing any locking parameters of the PID controller used in the previous experiments. The corresponding RF beat notes with modulation off and on are depicted in Fig. 5 (a,b) for comparison, respectively. When the modulation is on, the shape of the RF beat note is modified to the typical modulated pattern with two peaks located at the extremes of the modulation. Note that here we consider the idler SH, thus the modulation amplitude is doubled and results in ±4 MHz (see Fig. 5 (b)). When modulated, the central frequency of the RF beat note is precisely maintained, which would not be possible without the tight phase-locking. The $CEO_i$ modulation experiment is a good demonstration of the system's versatility. In practice this can be used in experiments where a dynamic control of the CEO is required. For instance, one could lock the comb to an external cavity for cavity-enhanced spectroscopy, in which case both the repetition rate and CEO often need to be precisely controlled[52]. Importantly, in our method these two parameters are independently adjustable, which is not necessarily the case with MLLs.

**Idler linewidth and intensity noise.** With regard to spectroscopy applications, there are two additional parameters of great importance – the comb tooth linewidth and the RIN. First, we estimated the comb tooth linewidth by another RF beat note measurement using a MIR CW HeNe laser (Research Electro-Optics, Model 30545) at 3390 nm. We do not have the exact information about the HeNe laser linewidth, thus this measurement only sets an upper limit for the comb tooth linewidth. The result can be seen in Fig. 6 (a) that shows an RF beat note with a 177 kHz FWHM (note the linear scale in Fig. 6 (a)). This is a narrow linewidth considering the fact that our pump MLL repetition rate is locked to an RF source, not to an ultra-stable CW-laser. We would like to emphasize that all the RF beat notes shown in Fig. 4 (e), 5 (a) and 6 (a) are detectable only if the OPG is seeded by the CW source. Without CW seeding, the $CEO_i$ is random and no RF beat note can be detected. See "Supplementary note 3: Pulse-to-pulse coherence" for more information.

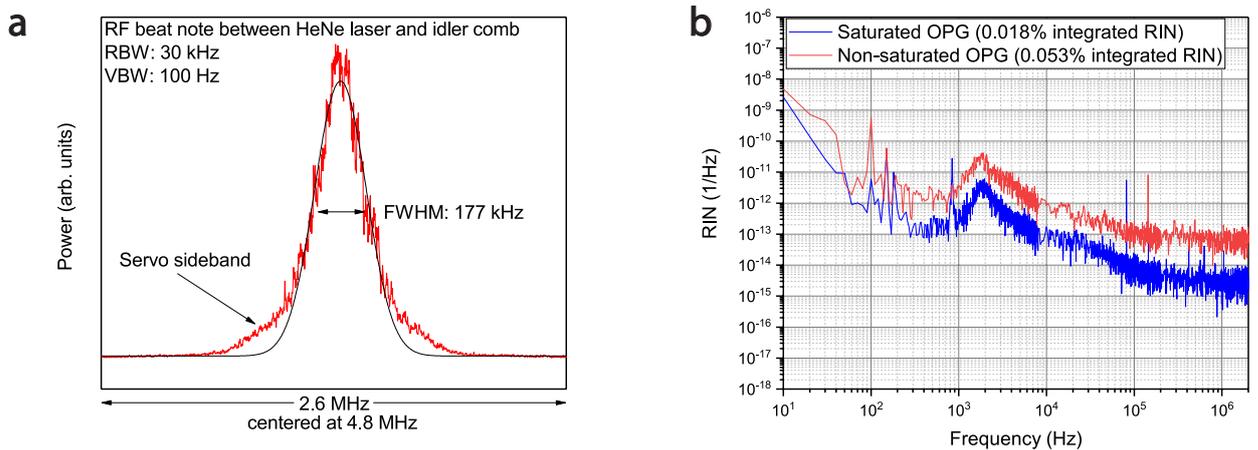

**Fig. 6 Idler linewidth and RIN.** (a) RF beat note between MIR HeNe laser and idler comb (red) (note the linear scale) and a Gaussian fit to the curve with servo sidebands excluded (black). (b) RIN of the idler comb in the saturated regime (blue) and non-saturated regime (red).

The RIN measurements were performed using the reference point with two different input powers that correspond to saturated and non-saturated OPG regimes (Fig. 6 (b)). At saturation we used an input pump power of 4.2 W that resulted in 450 mW of average output idler power. For the non-saturated regime, we used 3.2 W of average pump power that produced 170 mW of average idler power. The integrated RIN at the saturation is lower than for the non-saturated regime; we determined it to be 0.018% and 0.053% (integrated from 10 Hz to 2 MHz) for saturated and non-saturated regimes, respectively. The result is in agreement with the theoretical predictions as well as experimental observations by another research group[53].

## V. Conclusions

We have demonstrated a simple method that allows for the generation of stabilized MIR OFCs using femtosecond OPG with CW seeding. The CW seed laser plays a key role in the method, since it establishes the idler CEO stability. Moreover, the CW seed laser makes the system highly versatile allowing one to dynamically tune and modulate the idler CEO directly in the OPG process. Our setup does not have any cavities, which means that the repetition rate can also be freely tuned. We performed all the measurements required to prove the idler CEO stability, hence the usually challenging step of CEO determination can be omitted. An additional benefit of the scheme is its inherent insensitivity to relative timing jitter between the pump and seed, since the seed source is not pulsed. This is in contrast with DFG that relies on downshifted Raman solitons and requires a careful control of the temporal overlap between pump and signal pulses for intensity noise minimization[54]. Along with the already mentioned DFG and SPOPO methods, there are some examples of MIR OFCs that have been produced by spectral broadening in $Si_3N_4$[55, 56] and $LiNbO_3$[57] waveguides. These systems show a great promise in the miniaturization of the MIR combs, but unfortunately the output powers are quite limited in the MIR region. However, combining our method with nonlinear waveguide technologies could open up a new path to efficient and integrated MIR combs with dynamic CEO control.

**Supplementary material**

See supplementary material for details on the supercontinuum generation, phase-locking procedure and pulse-to-pulse coherence of the described OPG light source.

**Acknowledgements**

The work was funded by the Academy of Finland (Grant 314363).

**Data availability**

The data that support the plots within this paper and other findings of this study
are available from the corresponding author upon reasonable request.

## Supplementary Material for:

## Simple method for mid-infrared optical frequency comb generation with dynamic offset frequency tuning

Mikhail Roiz, Krishna Kumar, Juho Karhu & Markku Vainio

### Supplementary note 1: Supercontinuum generation

While preparing the experimental setup described in the main text, we tested two Photonic Crystal Fibers (PCF) for supercontinuum (SC) generation – NL-PM-750 and SC-5.0-1040-PM (NKT Photonics). NL-PM-750 resulted in the SC spectrum that can be seen in Supplementary Figure 1 (a) (also in Fig. 1b in the main text). This fiber has an extremely small mode field diameter of 1.6 µm and two Zero Dispersion Wavelengths (ZDW) located at 750 ± 15 nm and 1270 ± 30 nm. Due to the high nonlinearity and small mode field diameter, NL-PM-750 required a low average pump power of only 200 mW with ~50% coupling efficiency (meaning that only 100 mW in the fiber) to reach the desired wavelength range 1510 – 1560 nm. In fact, this fiber is able to cover the whole range of working wavelengths (1410 – 1560 nm signal seeding) of the Continuous-Wave (CW) seeded Optical Parametric Generation (OPG) described in the main text. Note that the coupling efficiency was measured using 10 mW of input power, so that the nonlinear processes would not contribute to the measurement. This fiber has anomalous dispersion region between the two zero dispersion wavelengths, and normal dispersion everywhere else. Thus, the long wavelength part of the spectrum (>1300 nm) can be attributed to the so-called rogue waves[1], rather than downshifted Raman solitons[2]. In this case, the Raman soliton is located at 1230 nm, which corresponds to the anomalous dispersion region as expected.

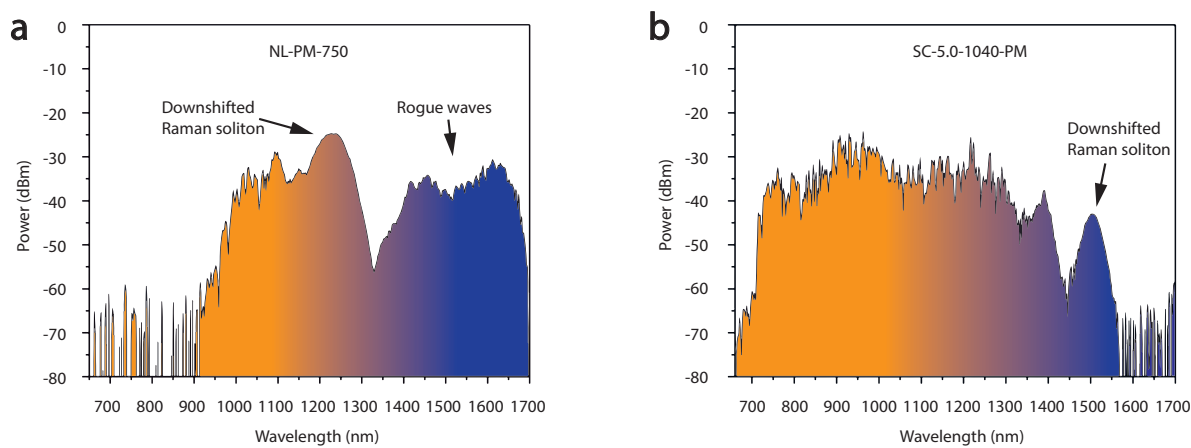

**Supplementary Figure 1. SC generation.** (a) SC spectrum generated using NL-PM-750 (80 cm long) with 200 mW of average pump power. (b) SC spectrum generated using SC-5.0-1040-PM (190 cm long) with 1.5 W of average pump power.

In the main text we demonstrated that NL-PM-750 fiber leads to an excellent performance of the idler Carrier-Envelope Offset ($CEO_i$) frequency stabilization when the CW seed laser is phase-locked to the generated SC. Nevertheless, the same performance in terms of $CEO_i$ stabilization can be expected with the second fiber - SC-5.0-1040-PM (1040 ±15 µm single ZDW), where the downshifted Raman soliton has the longest wavelength instead of the rogue waves (see Supplementary Figure 1 (b)). Unfortunately, the SC generation in SC-5.0-1040-PM is not as efficient at long wavelengths as in the case of NL-PM-750. It requires a high average pump power of 1.5 W (~50% coupling efficiency, so 0.75 W reaches the fiber) and the peak of the Raman soliton does not even reach the 1510 nm (the shortest wavelength available in our CW seed laser). Further increase in the average pump power did not lead to any changes of the Raman soliton peak.

However, we were able to phase-lock the CW seed laser at 1510 nm using the long wavelength side of the Raman soliton and do the stability measurements.

As a conclusion, here we demonstrated that not only the conventional downshifted Raman solitons can be used for the $CEO_i$ stabilization in our setup, but also the rogue waves. In addition, other alternative ways of SC generation exist, for instance those based on second-order nonlinearities in $LiNbO_3$ waveguides[3]. Indeed, the possibility of using different approaches for SC generation is another advantage of our method over the offset free Difference Frequency Generation setup mentioned in the main text that relies on the transform limited Raman solitons for the second input to work[4].

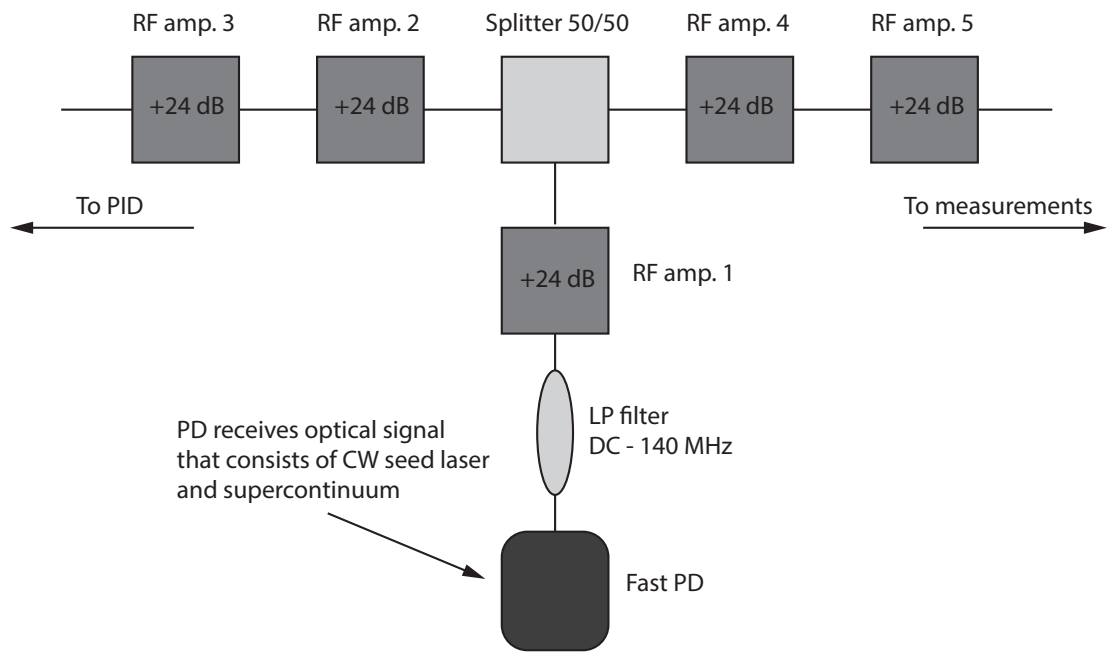

**Supplementary Figure 2. Phase-locking RF circuit.** Fast PD is used to detect the RF beat note between the CW seed laser and the SC.

## Supplementary note 2: Phase-locking

In this supplementary note we discuss the phase-locking procedure mentioned in the main text. The corresponding Radio Frequency (RF) circuit for the phase-locking is shown in Supplementary Figure 2. We used a fiber-coupled fast photodiode (Thorlabs, DET08CFCM) to detect the RF beat note between the CW seed laser and the SC. The generated RF signal from the photodiode is low-pass filtered in order to get rid of the repetition rate RF harmonics (250 MHz, 500 MHz, 750 MHz etc.). Note that the low-pass filter sets an upper limit for the RF frequency in the phase-locking process, and thus it also limits the $CEO_i$ tuning range. Next, the RF beat note signal is amplified using RF amplifiers (Mini-Circuits, ZFL-500LN+) and split into two arms: one for the phase-locking process and the other for measurements including beat note, phase-noise and frequency counting (see Fig. 4a-d in the main text). The RF beat note is needed to produce an error signal for the phase-locking of the CW seed laser. The error signal is produced using a frequency mixer where an RF Local Oscillator (LO) output is mixed with the detected RF beat note. Our Proportional–Integral–Derivative (PID) controller (Toptica Photonics, mFALC 110) has an inbuilt frequency mixer; this is why the frequency mixer is not present in Supplementary Figure 2. The PID controller used in this work has two output channels, one for fast feedback and another for slow corrections. The fast channel is connected directly to the laser chip and controls the laser frequency via current (bias-T). The slow channel is needed for compensating slow drifts. It is connected to the control unit of the laser, which adjusts the laser frequency via piezoelectric actuators.

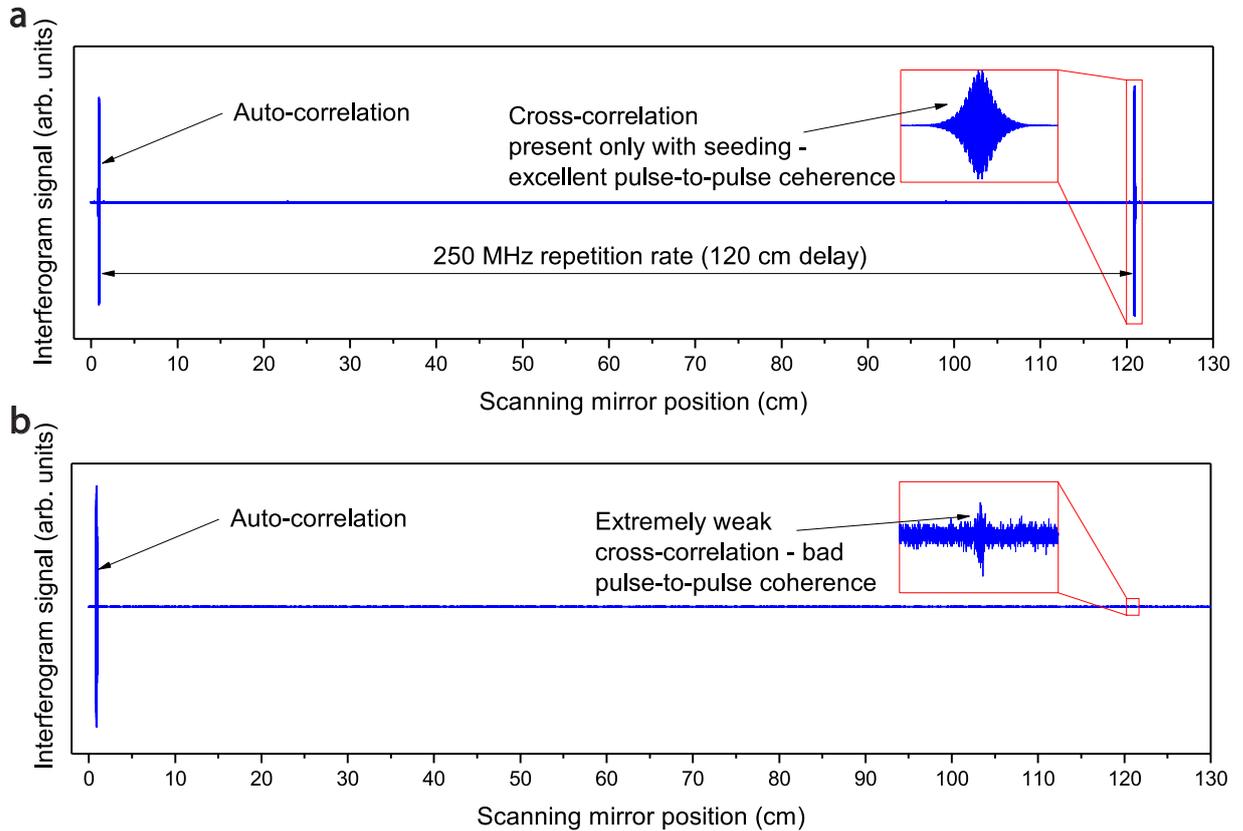

**Supplementary Figure 3. Pulse-to-pulse coherence measurement of the signal comb.** (a) Interferogram of the signal comb when the CW seeding is enabled; note that both the auto- and cross-correlation are present. (b) Interferogram of the signal generated in the OPG without CW seeding; note that the cross-correlation is extremely weak indicating lack of the pulse-to-pulse coherence.

## Supplementary note 3: Pulse-to-pulse coherence

In this supplementary note we demonstrate pulse-to-pulse coherence of the OPG setup described in the main text. For technical reasons, we did this measurement only for the signal comb at the reference point (1530 nm signal and 3400 nm idler). We used a Fourier transform infrared spectrometer (Bruker, IFS 120 HR) as a Michelson interferometer with a very long scanning arm. In order to characterize the pulse-to-pulse coherence, we intend to resolve both the auto- and cross-correlations of the generated signal pulses on the interferogram. Auto-correlation represents the interference of the pulse with itself. Obviously, the pulses under consideration are perfectly coherent with themselves, thus the auto-correlation is expected to be present in both cases when the CW seeding is enabled (see Supplementary Figure 3 (a)) and without CW seeding (see Supplementary Figure 3 (b)). On the other hand, cross-correlation shows coherence of two subsequent pulses in the pulse train averaged over many pulses, thus it is indicative of pulse-to-pulse coherence. Indeed, as evident from Supplementary Figure 3 (a), the strong cross-correlation is present (with the delay of 120 cm corresponding to 250 MHz repetition rate) in the case of CW seeding indicating excellent pulse-to-pulse coherence. However, when the OPG is not seeded, the CEO of the signal pulses is random, which means that even two subsequent pulses can barely interfere with one another (see Supplementary Figure 3 (b)).